\documentclass[fleqn,twoside]{article}
\usepackage{espcrc2}

\title{The Universality of Seesaws}

\author{R. D. Peccei
\address{Department of Physics and Astronomy\\
University of California at Los Angeles\\
Los Angeles, California, 90095}}

\begin{document}
\begin{abstract}

I discuss how the ideas associated with Seesaws, first introduced in the context of neutrino masses, are generally useful for understanding the very disparate scales one encounters in particle physics. From this point of view, the energy scale characterizing the Universe's dark energy presents a real challenge. A natural Seesaw explanation for this scale ensues if one imagines tying the dark energy sector to the neutrino sector, but this idea requires bold new
dynamics.

\vspace{1pc}

\end{abstract}

\maketitle

\section{A PLETHORA OF SCALES}

One of the most difficult problems to understand in particle physics is
the disparity of the mass scales that characterize fundamental interactions.
A prototypical example is provided by the large hierarchy between the
physical scale associated with gravitational interactions (the Planck
mass, $M_P = G_N^{-1/2} =1. 22 \times 10^{19}$ GeV) and that connected
with the electroweak interactions (the Fermi scale,
$v_F=(\sqrt{2}G_F)^{-1/2} \simeq 250$ GeV): 
\begin{equation}
M_P/v_F \simeq 10^{17}.
\end{equation}
Another challenging issue is provided by the fine structure in the mass
spectrum of quarks and charged leptons:
\begin{equation}
m_q = \{\rm{ 5 ~MeV -175 ~GeV}\}
\end{equation} 
\begin{equation}
 m_\ell=\{\rm{0.5~MeV-2~GeV}\},
\end{equation}
whose origin remains a mystery. A third example is provided by the very
small mass scales associated with neutrinos, compared to that of quarks
and charged leptons:
\begin{equation}
m_{\nu}=\{\rm{4\times 10^{-3} eV- 2~eV}\}. 
\end{equation}

In this last example, however, it is possible to understand the general
magnitude of $m_{\nu}$ from the Seesaw mechanism \cite{GMRSY}, whose
twenty-fifth  anniversary we are celebrating in this Symposium. As a
result of  the Seesaw mechanism, neutrino masses are small because they
reflect the presence of a much larger physical scale in the theory.
Thus, typically, 
\begin{equation}
m_{\nu} \sim v_F^2/ M_N;~~{\rm{or}}~~ m_{\nu} \sim v_F^2/ M_X,
\end{equation}
with ${M_N, M_X}$ associated with either a very heavy right-handed
neutrino $ M_N\sim 10^{11}-10^{15}\rm{GeV}$ or the GUT scale $
M_X\sim 10^{15}-10^{16}\rm{GeV}$.

When thinking about the issue of scales in particle physics,
traditionally one takes the Planck mass $M_P$ as input and one asks
questions about the origin of the light scales. There is a real plethora
of such light scales, some arising from experimental input while others
being pure theoretical constructs. I display some of these scales in
Table 1. As one can see from this Table, these scales range over  30
orders of magnitude! Interrelating these scales is a real challenge and
requires making assumptions on physics beyond the Standard Model. What I
will try to argue here is that the existence of Seesaws may provide a
useful guiding principle to help sort out what is fundamental and what
is derivable among this panoply of scales.

\begin{table*}[htb]
\caption{A sample of particle physics scales}
\label{ Table 1}
\begin{tabular}{ lll}
\hline 
Scale & Physics & Value (GeV)\\
\hline
$M_P$ & Gravity & $1.2 \times 10^{19}$\\
$M_X$ & GUTS & $2 \times 10^{16}$\\
$M_N$ & RH neutrino & $10^{11}-10^{15}$\\
$f_{PQ}$ & PQ breaking & $10^{9}-10^{12}$\\
$\mu_S$ & SUSY breaking & $10^{5}-10^{15}$\\
$v_F$ & EW breaking & 250\\
$M_H$ & Higgs & $<$ 180\\
$\Lambda_{QCD}$ & QCD & 0.3\\
$m_q$ & quarks &  0.005-175\\
$m_{\ell}$ & leptons &  $5\times 10^{-4}$- 2\\
$m_{\nu}$ & neutrinos & $10^{-12}-10^{-9}$\\
\hline
\end{tabular}\\[2pt]
\end{table*}

The only scale in Table 1 which has a theoretically pristine origin is
$\Lambda_{QCD}$, since this scale is set by the strong QCD dynamics
itself. Roughly speaking, one can define $\Lambda_{QCD}$ as the scale
where the QCD coupling constant becomes strong:
$\alpha_s(\Lambda_{QCD}^2)=1$. The relation of $\Lambda_{QCD}$ to $M_P$
is logarithmic and the only question really is why
$\alpha_s(M_P^2)\simeq 1/45$. Could this be a boundary condition coming
from physics at the Planck scale? Furthermore, because QCD is a
dynamical theory, there is a close correlation between the physical
scale $\Lambda_{QCD}$ and the masses of the physical hadronic states.
Indeed, $M_{\rm{hadrons}} \sim \Lambda_{QCD}$.{\footnote{Pions are an
exception because of their Nambu-Goldstone nature, with $ m_{\pi}^2
\sim m_q \Lambda_{QCD}$.}}

The situation is much different in the electroweak theory. First, it is
unlikely that the Fermi scale $v_F$ is a dynamical scale, just like
$\Lambda_{QCD}$, since precision electroweak experiments favor a light
Higgs and disfavor QCD-like Technicolor theories. \cite{EW} Second,
although $m_q$ and $m_{\ell}$ are proportional to $v_F$, the fact that the mass
spectrum of quarks and charged leptons spans five orders of magnitude
suggests that the Yukawa couplings which engender these masses arise
from physics at scales much larger than $v_F$. Third, and this is the
real problem,  in the Standard Model the sensitivity of $v_F$ to any
high energy cut-off is quadratic, so that the hierarchy $ M_P/v_F \sim
10^{17}$ is very hard to understand.

I do not believe this hierarchy problem is resolved by
extra-dimensional theories, \cite{d>4} where one assumes that the Planck
scale in (d+4)-dimensions is the Fermi scale: $M^d_P\simeq v_F$. These
theories involve introducing a compactification radius R, whose scale is
set by demanding that in 4-dimensions the scale of gravity is $M_P$.
This requires that
\begin{equation}
M_P = M^d_P(M^d_P R)^{d/2} \simeq v_F (v_F R)^{d/2}.
\end{equation}
Obviously, the hierarchy problem $M_P/v_F\sim 10^{17}$ is replaced in
these theories by the question of why $v_F R>>1$.

\section{SEESAWS AS DYNAMICAL SOLUTIONS}

In my view, it is much more satisfactory to think of $v_F$ as
originating from a Seesaw. This obtains in supersymmetric (SUSY) theories
spontaneously broken in a hidden sector which is coupled to matter by gravity
mediated interactions. \cite{sugra} First of all, as is well known, the
presence of a supersymmetry modifies the dependence of $v_F$ on a
high-energy cut-off from quadratic to logarithmic. Second, if indeed
SUSY is spontaneously broken at a scale
$\mu_S$ in a hidden sector coupled to matter only gravitationally, the
superpartner masses (and other SUSY breaking parameters) are naturally
given by a Seesaw formula:
\begin{equation}
\tilde{m} \simeq \mu_S^2 /M_P.
\end{equation}
However, because of the large top Yukawa coupling, in this scenario
\cite{IR} the breaking of supersymmetry naturally engenders a
concomitant breaking of the electroweak theory. What happens,
essentially, is that positive mass squared parameters in the Higgs
sector at the SUSY breaking scale $\mu_S$ evolve at scales of $O(v_F)$
to negative mass squared parameters. That is,
\begin{equation}
\mu^2(\mu_S^2) \rightarrow -\mu^2(v_F^2).
\end{equation}
Since $ \mu^2(\mu_S^2) \sim \tilde{m}^2 $, it follows that $v_F$ is also
given by a Seesaw formula:
\begin{equation}
v_F \simeq \mu_S^2 /M_P.
\end{equation}

For neutrinos, via the Seesaw mechanism, one obtains an understanding of
the small scale for neutrino masses from the presence of a large new
scale (either $M_N$ or $M_X$) and a known intermediate scale $v_F$. If,
indeed, the origin of the Fermi scale is due to the above SUSY induced
Seesaw, we have effectively tied this scale $v_F \simeq $ 250 GeV to a
much larger scale $\mu_S \sim 10^{11}$ GeV. In this Seesaw, in contrast
to the neutrino case, we have used {\bf{known}} low and high scales
($v_F$ and $M_P$, respectively) to infer the existence of an
intermediate scale $\mu_S$.

From a fundamental point of view, there might appear to be no real
advantage to having "explained" $v_F$  through the introduction of a new
intermediate scale $\mu_S$, except to have shortened the gap between the
Planck scale and the driving scale $\mu_S$ for the physics we observe.
The hierarchy to explain now is not $M_P/v_F \sim 10^{17}$ but
$M_P/\mu_S \sim 10^{8}$. However, in the literature, many attempts exist
to use the scale $\mu_S$ also as the prototypical scale where family
structure originates. \cite{family} Typically, this is done by
exploiting variants of the Froggatt-Nielsen \cite{FN} mechanism, with
small Yukawa couplings being given by VEV ratios, $\Gamma \sim
[<\sigma>/ \mu_S]^n$, with the VEV $<\sigma>$ breaking some assumed
family symmetry.

I will not pursue these matters further here, except to remark that one
can systematically relate most of the scales that enter in Table 1 to
physics at higher scales via some kind of Seesaw formulas. {\footnote{
As an example, for instance, the axion mass is given by the Seesaw
formula $m_a \sim m_q^{1/2}\Lambda^{3/2}_{QCD} / f_{PQ}$ involving the
ratio of the scale where the axial anomaly becomes relevant and the
scale where $U(1)_{PQ}$ breaks down spontaneously. The presence of the
quark mass in the formula for $m_a$ reflects the influence of additional
chiral symmetries in QCD in the massless quark limit.}} Instead I want
to spend some time discussing how these ideas of having Seesaw formulas
tie low energy parameters to physics at higher scales runs into a
significant challenge when one tries to address the issue of dark energy
in the Universe.

\section{DIALING SCALES THROUGH THE UNIVERSE}

Einstein's equations describing the expansion of the Universe in a
Robertson-Walker background provide a wonderful scale-meter. The Hubble
parameter at different temperatures during the expansion provides the
yardstick. Although now $ H_o = (1.5 \pm 0.1) 10^{-33}$ eV is a tiny
scale, since the value of the Hubble parameter varies with temperature
as the Universe cools it samples all scales, from the Planck mass
downwards.

Einstein's equations, written in terms of the Robertson-Walker scale
factor $R$
\begin{equation}
H^2 \equiv (\frac {\dot{R}}{R})^2= \frac{8\pi G_N \rho}{3}
-\frac{k}{R^2} + \frac{\Lambda}{3} 
\end{equation}
\begin{equation}
\frac{\ddot{R}}{R} = \frac{\Lambda}{3} -\frac{4 \pi G_N}{3}(\rho
+3p)
\end{equation}
determine $H$ and the Universe's acceleration once $\rho$, $p$, $k$ and
$\Lambda$ are specified. In a flat Universe [$k=0$], as predicted by
inflation \cite{inf} and confirmed observationally by WMAP, \cite{WMAP}
the Universe's expansion accelerates if  the cosmological constant is
such that $\Lambda > 4 \pi G_N \rho_{\rm{matter}}$. Alternatively, if
$\Lambda=0$, the expansion of the Universe accelerates if a dominant
component of the Universe has negative pressure and $\rho +3p <0$. The
observed acceleration is evidence for this dark energy.

It is convenient in what follows to set the parameter $\Lambda=0$ adding, however, a
dark energy contribution explicitly to the density. That is, $\rho
\rightarrow \rho +\rho_{\rm{dark~ energy}}$. It is easy to see then that the
pure cosmological constant case corresponds to assuming the equation of
state 
\begin{equation}
\omega_{{\rm dark~ energy}}= \frac{p_{{\rm dark~ energy}}}{\rho_{{\rm dark ~energy}}}= -1,
\end{equation}
where $\rho_{{\rm dark~ energy}}= \rho_{{\rm vacuum}}= \rm{constant}$. Writing
\begin{equation}
H^2= \frac{ 8 \pi G_N \rho}{3} +\frac{ 8 \pi G_N \rho_{{\rm dark~ energy}}}{3}
\end{equation}
we know observationally that, at the present time, $H_o^2$ gets about
30\% contribution from the first term and 70\% from the second. So there
appear to be two Seesaws to explain:
\begin{equation}
H_o \simeq \frac{\rho_o^{1/2}}{M_P}~;~\rm{and}~~ H_o \simeq \frac{\rho_{{\rm dark
~energy}}^{1/2}}{M_P}.
\end{equation}
The first Seesaw, which in fact is not a true Seesaw, is understood in
terms of known, or at least speculated, physics. The other Seesaw,
involving the dark energy density, is totally mysterious.

Let me try to address both points below. Because the energy density
depends on the Universe's scale factor $R$ as 
\begin{equation}
\rho \sim R^{-3(1+\omega)},
\end{equation}
the contribution of $\rho_{{\rm dark~energy}}$ to $H^2$ at earlier times is
negligible, so that
\begin{equation}
H^2 \simeq \frac{8 \pi G_N \rho}{3}.
\end{equation}
Because the Hubble parameter depends on temperature, $H=H(T)$, the above
is really a dynamical equation, not a Seesaw. The total density $\rho$
just fixes the rate of the Universe's expansion.

As the Universe expands different components dominate $\rho$, as they,
in general, have different temperature dependences and different
threshold factors. Schematically, one can write
\begin{equation}
\rho =\rho_{{\rm radiation}} +\rho_{{\rm matter}} +\rho_{{\rm dark~ matter}},
\end{equation}
with
\begin{equation}
\rho_{{\rm radiation}}=[\frac{\pi^2}{30}]g(T) T^4
\end{equation}
\begin{equation}
\rho_{{\rm matter}}=[\frac{2\xi(3)}{\pi^2}]\{M_B\eta +\Sigma_i m_{\nu_i}\}
T^3
\end{equation}
\begin{equation}
\rho_{{\rm dark~ matter}}\simeq \{ f_{PQ} \Lambda_{QCD}
+\frac{m^*}{T^* <\sigma {\rm{v}}>}\} \frac{T^3}{M_P}.
\end{equation}

In the above, the different components in $\rho$ become effective when
the Universe cools below the relevant decoupling temperature. For
instance, baryons contribute after nucleosynthesis starts (
$T_{{\rm decoupling}} \sim$ MeV); neutrinos turn on below $T \simeq m_{\nu_i}$;
axionic dark matter, if it exists, turns on below the QCD phase
transition; and neutralino dark matter, if it exists, turns on below the
electroweak phase transition. At any rate, at present [$T_o \simeq
3^{o} K;  g(T_o)=2$]  the contribution of $\rho_{{\rm radiation}}$ is negligible, while the
particle physics parameters {$M_B,\eta, m_{\nu_i}, f_{PQ}$, etc} insure
that $\rho_{\rm {matter}}$ and $\rho_{{\rm dark~ matter}}$ contribute, respectively,
approximately 2\% and 28\% to $H_o^2$.

The situation is quite different with the second Seesaw. Here, for
example, if the dark energy is due to the presence of a cosmological
constant, so that $\rho_{{\rm dark~ energy}}= \rho_{{\rm vacuum}}= E_o^4$, one has a
real Seesaw:
\begin{equation}
H_o \simeq \frac{E_o^2}{M_P},
\end{equation}
which  fixes $E_o \simeq 2 \times 10^{-3}$ eV. What physics is
associated with this small energy scale? All vacuum energies we know in
particle physics are enormously bigger. For instance, $E_o^{QCD} \sim
\Lambda_{QCD} \sim 1$ GeV.

Matters are not substantially different if $\rho_{{\rm dark~ energy}}$ has a
more dynamical origin. Although now the dark energy is temperature
dependent
\begin{equation}
\rho_{{\rm dark ~energy}}=\rho_{{\rm dark~ energy}}^o[T/T_o]^{3(1+\omega)},
\end{equation}
the parameters characterizing the dynamical theory are very difficult to
understand. Let me illustrate this point by considering quintessence
\cite{quint} as a model for the dark energy of the Universe.

 In this
case one associates the dark energy with a new scalar field $\phi$ which
has negative pressure. In the present epoch we know that $ \rho_{{\rm dark
~energy}}\simeq 0.7 [3H_o^2/ 8 \pi G_N] $ and that the dark energy
equation of state gives a range for $\omega_{{\rm dark~ energy}}$ between -0.7
and -1.2. \cite{omega}. Thus, if quintessence is the source for the dark
energy, we must have that
\begin{equation}
\rho_{{\rm quint}}=\frac{\dot{\phi}^2}{2 }+V(\phi) \simeq 0.7 [\frac{3H_o^2}{
8 \pi G_N}]   
\end{equation}
\begin{equation}
p_{{\rm quint}}=\frac{\dot{\phi}^2}{2 }-V(\phi) \simeq 0.7\omega
[\frac{3H_o^2}{ 8 \pi G_N}].     
\end{equation}

The field $\phi$ is dynamical and to realize the above equations the
magnitude of the field $\phi$ must be large, of order of the Planck mass
itself: $\phi \sim M_P$. With such large fields it is impossible to
obtain the above results for $\rho$ and $p$ unless the field $\phi$ has
nearly zero mass:
\begin{equation}
m_{\phi} \sim \frac{E_o^2}{\phi} \sim H_o \simeq 10^{-33} \rm{eV}.
\end{equation}
The above Seesaw formula, however, is unprotected from getting big mass
shifts, unless the quintessence field essentially decouples from matter.
\cite{decoup}

\section{NEUTRINOS TO THE RESCUE?}

In a sense, the quintessence interpretation of $\rho_{{\rm dark~ energy}}$
results in a very unpalatable Seesaw:
\begin{equation}
m_{\phi} \simeq \frac{E_o^2}{M_P},
\end{equation}
where a difficult to understand scale $E_o \sim 2\times 10^{-3}$ eV
produces, from a particle physics point of view, an even more difficult
to understand scale $m_{\phi} \sim H_o \simeq 10^{-33} $ eV. It would be
much more satisfactory if one could understand the dark energy density
as arising dynamically from a known particle physics scale.

A very interesting suggestion along these lines has been put forward
recently by Fardon, Nelson and Weiner.\cite{FNW} Its starting premise
is that one should be able to explain in a natural fashion why, in the
present epoch, the energy density associated with dark energy and matter
should be approximately the same: $ \rho_{{\rm dark~ energy}} \simeq
\rho_{{\rm matter}}$. Because these densities, presumably, have different
temperature dependences, their near equality now itself is a mystery. As
many people have noted, this coincidence is resolved dynamically if the
dark energy density tracks (some component) of the matter density.
\cite{tracking} What Fardon, Nelson and Weiner suggest is that
$\rho_{{\rm dark~ energy}}$  tracks the energy density in neutrinos,
$\rho_{\nu}$. This avoids some of the issues that would arise if
$\rho_{{\rm dark~ energy}}$ were really to track some better known component of
$\rho_{{\rm matter}}$, like $\rho_{B}$.  Further it perhaps allows one to
understand the scale $E_o$ associated with $\rho_{{\rm dark~ energy}}$ in terms
of the scale of neutrino masses, which are of a similar magnitude.

The idea put forth by Fardon, Nelson and Weiner [FNW] \cite{FNW} is
quite radical. By imagining that the neutrinos and the dark energy are
coupled together, the energy density associated with the dark energy
depends on the neutrino masses. In turn, these masses are not fixed but
are variable, with their magnitude being a function of the neutrino
density, $m_{\nu}=m_{\nu}(n_{\nu})$. Assuming for simplicity just one
flavor of neutrino, in the FNW picture the energy density in the dark
sector (neutrinos plus dark energy) is given by:
\begin{equation}
\rho_{{\rm dark}}=m_{\nu}n_{\nu} +\rho_{{\rm dark~ energy}}(m_{\nu}).
\end{equation}
This energy density will stabilize when
\begin{equation}
n_{\nu} +\rho^{\prime}_{{\rm dark~ energy}}(m_{\nu}) =0.
\end{equation}

The above equations have an immediate implication for the equation of
state in the dark sector. Since
\begin{equation}
\omega +1=-\frac{\partial \ln \rho_{{\rm dark}}}{3 \partial \ln R}, 
\end{equation}
it follows that
\begin{eqnarray}
\omega +1= -[\frac{R}{3\rho_{{\rm dark}}}] \{m_{\nu}\frac{\partial n_{\nu}}{\partial
R} + n_{\nu}\frac{\partial m_{\nu}}{\partial R} \nonumber\\
 +\rho^{\prime}_{{\rm dark
~energy}}\frac{\partial m_{\nu}}{\partial R}\}, 
\end{eqnarray}
whence
\begin{equation}
\omega +1=\frac {m_{\nu}n_{\nu}}{\rho_{{\rm dark}}}= \frac{m_{\nu}n_{\nu}}{
m_{\nu}n_{\nu}+\rho_{{\rm dark ~energy}}}.
\end{equation}

We see from this equation that if $\omega \simeq -1$, the neutrino
contribution to $\rho_{{\rm dark}}$ is a small fraction of $\rho_{{\rm dark
~energy}}$. Further, since we expect that $\rho_{{\rm dark ~energy}} \sim R^{-
3(1+\omega)}$, if $\omega$ does not change significantly with $R$, it
follows that the neutrino mass is nearly {\bf inversely proportional} to
the neutrino density:
\begin{equation}
m_{\nu} \sim n_{\nu}^{\omega}.
\end{equation}

I will not discuss the FNW scenario much further here, but will make
just a few remarks:

\begin{enumerate}
\item                    If $\omega $ is near its central value, $\omega = -
0.8$, then the dark sector equation of state predicts that in the
present epoch $m_{\nu}^{{\rm cosmo}} \simeq 5$ eV. However, since $ m_{\nu}
\sim n_{\nu}^{\omega}$, if there is neutrino clustering in our galaxy,
the observed neutrino mass on earth could be much smaller: $
m_{\nu}^{{\rm obs}} \simeq 5[ n_{\nu}^{{\rm local}}/ n_{\nu}^{{\rm cosmo}}]^{\omega}$ eV.

\item                  The variability of neutrino masses with their
density requires re-examining many of the astrophysical and cosmological
constraints on neutrinos coming from big bang nucleosynthesis,
supernovas, leptogenesis, etc. Remarkably, the FNW scenario seems to
survive rather unscathed by these constraints. \cite{FNW} \cite{others}

\item                Although the detailed dynamics of the dark sector is
unclear at this stage, it is most likely that the coupling between dark
energy and neutrinos comes through the $SU(2) \times U(1)$ singlet
$M_N$. So, the neutrino masses feel the field responsible for the dark
sector dynamics,  $\phi_{{\rm dark~ energy}}$, via the Seesaw: $m_{\nu} \sim
v_F^2/M_N(\phi_{{\rm dark~ energy}})$.

\item                This scenario naturally "explains" why the energy
scale associated with the dark energy, $E_o$, is of the same order as
that of neutrino masses, since these components of the Universe track
each other. Indeed, $E_o$ is set by the same Seesaw which gives neutrino
masses: $E_o \sim m_{\nu} \sim v_F^2/M_N(\phi_{{\rm dark~ energy}})$.

\end{enumerate}

\section{CONCLUDING REMARKS}

I hope that the discussion presented above has helped to show that the
ideas associated with Seesaws are useful when one wants to reach some
understanding of the very disparate scales one encounters in particle
physics.
In fact, most if not all of the scales entering particle physics can be
associated, in one form or another, to a Seesaw. So, in a sense, Seesaws
provide a universal framework to think about disparate scales.

From this point of view, the dark energy scale $E_o \sim 2 \times
10^{-3}$ eV presents a real challenge. Straightforward particle physics
models of dark energy, like quintessence, \cite{quint} involve even less
understandable mass scales, which cannot be protected unless
quintessence decouples from everything else.\cite{decoup} The FNW
scenario just discussed \cite{FNW} allows a natural Seesaw explanation
for $E_o$. However, the idea of tying the dark energy sector to the
neutrino sector is very speculative and requires imagining bold new
dynamics.

\section*{ACKNOWLEDGEMENTS}

I am grateful to the Fujihara Foundation and Yoji Totsuka and  Kenzo Nakamura for the spendid hospitality provided at the Seesaw 1979-2004 Seminar. This work has been supported in part by the Department of Energy under Contract No. FG03-91ER40662, Task C.

\end{document}